# A Learnable Despeckling Framework for Optical Coherence Tomography Images

Saba Adabi,[a,b] Elaheh Rashedi,[c] Hamed Mohebbi-Kalkhoran,[a] Xue-wen Chen,[c] Silvia Conforto,[a] and Mohammadreza Nasiriavanaki,[b,d]

[a] Applied Electronics Department, Roma Tre University, Rome, Italy
[b] Biomedical Engineering Department, Wayne State University, Detroit, MI, USA
[c] Department of Computer Science, Wayne State University, Detroit, MI, USA
[d] Barbara Ann Karmanos Cancer Institute, Detroit, MI, USA

**Abstract**. Optical coherence tomography (OCT) is a prevalent, interferometric, high-resolution imaging method with broad biomedical applications. Nonetheless, OCT images suffer from an artifact, called speckle which degrades the image quality. Digital filters offer an opportunity for image improvement in clinical OCT devices where hardware modification to enhance images is expensive. To reduce speckle, a wide variety of digital filters have been proposed; selecting the most appropriate filter for each OCT image/image set is a challenging decision. To tackle this challenge, we propose an expandable learnable despeckling framework, we called LDF. LDF decides which speckle reduction algorithm is most effective on a given image by learning a figure of merit (FOM) as a single quantitative image assessment measure. The architecture of LDF includes two main parts: (i) an autoencoder neural network, (ii) filter classifier. The autoencoder learns the figure of merit based on the quality assessment measures obtained from the OCT image including signal-to-noise ratio (SNR), contrast-to-noise ratio (CNR), equivalent number of looks (ENL), edge preservation index (EPI) and mean structural similarity (MSSIM) index. Subsequently, the filter classifier identifies the most efficient filter from the following categories: (a) sliding window filters including median, mean, and symmetric nearest neighborhood; (b) adaptive statistical based filters including Wiener, homomorphic Lee, and Kuwahara; and, (c) edge preserved patch or pixel correlation based filters including non-local mean, total variation, and block matching 3D filtering.

## 1 Introduction

Optical coherence tomography (OCT) is a non-invasive, low coherent interferometry technique that uses backscattered light to generate tomographic images of tissue microstructures (1). With advances in imaging technology and optical devices, OCT currently has extensive biomedical applications in various fields including dermatology (2). Although, OCT offers high-resolution, three-dimensional (3D) images of tissues with 2 µm to 15 µm resolution and 1 mm to 2 mm imaging depth (3), OCT images suffer from a grainy texture artifact, speckle, due to the broadband / low coherent light source used in the configuration of OCT (3). Speckle is formed, if the out-of-phase backscattered signals reach the detector within the coherence time of the laser and it is the summation of multiple optical backscattered wavefields from the sample arm of the interferometer (4). Speckle reduces the image quality, e.g., spatial resolution of the borders and edges in the image (5). Since speckle is an artifact that carries sub-micron structural information, devising an appropriate speckle reduction algorithm becomes a challenging task (6). Speckle reduction methods can be categorized to hardware (compounding) methods(7) and software based methods (8, 9). In compounding methods, the sample is imaged several times by the OCT system and there are angular compounding where sample imaged from different angles; frequency compounding where sample imaged with different wavelengths; polarization compounding where sample imaged with different polarizations; and spatial compounding where sample imaged from different positions(4, 7, 9). Software-based methods or digital filters rely on a mathematical model of the speckle and are implemented based upon the local first-order or overall statistics of the image. Among speckle reduction digital filtering methods, the most common ones such as averaging and median are time efficient but reduce the spatial resolution. Improved performance is provided by adaptive digital filters including Lee (10), Kuwahara (11) and the Wiener filter (12). Moreover, Wavelet-based filtering methods (9) and a diffusion-based filter with fuzzy logic thresholding as transform domain techniques have demonstrated satisfactory results (13). Some of these techniques are compared in (8) and authors concluded that the combination of enhanced-Lee method and Wiener filter could significantly improve the quality of OCT images of *ex vivo* bovine retina. The other filters were developed based on an A-scan reconstruction procedure (14) or Bayesian estimation (15) for OCT image speckle reduction. Several other studies have been carried out on speckle reduction by tuning the internal parameters in a particular algorithm (16). Relying on more advanced filtering methods, the speckle reduction methods were developed based on the total variation (TV) concept (17), non-local mean (NLM) filtering (17), and block-matching 3D filtering (BM3D).

Mathematical models are used to describe the speckle pattern depends on scatterer's features. A tissue's cellular specification can be represented by optical and textural features extracted from OCT images. Together with the attenuation coefficients (18), textural features can provide valuable information about the texture and scattering properties of a biological tissue (19, 20). We introduced an expandable despeckling framework, named Learnable Despeckling Framework (LDF). LDF decides which speckle reduction algorithm is most effective for a given image by learning a single quantitative image assessment measure: the figure of merit (FOM). Among several algorithms developed for speckle reduction in OCT images, we utilized the above digital filters to evaluate our proposed framework. Any other algorithm can easily be added to this framework. LDF's architecture is composed of two main parts; an autoencoder neural network and a filter classifier. In our proposed scheme, initially the autoencoder learns to construct a weighted FOM measure using quality assessment measures, i.e., SNR, CNR, ENL, EPI and MSSIM, extracted from filtered OCT images in the following categories: (a) sliding window filters including median, mean, and SNN; (b) adaptive statistical based filters including Wiener, homomorphic Lee, and Kuwahara; and, (c) edge preserved patch or pixel correlation based filters including non-local mean, total variation, and BM3D. Then, the filter classifier identifies the most efficient filter from the mentioned categories.

## 2 Materials and Methods

In section 2.1, we present the digital filtering methods used in this study. In section 2.2, we introduce different features that can be extracted from the raw OCT images. Then in section 2.3, we explain the image quality assessment measures that are utilized to evaluate the quality of the filtered images. Following that, in section 2.4, we introduce architecture of our learnable despeckling framework in details. The OCT system specifications is then given in 2.5.

### 2.1 Digital Filtering Methods

Digital filters are means to implement mathematical operations characterized by their transfer functions; or equivalently, on a sampled signal or image, in order to improve their quality. Several speckle reduction methods have been designed and implemented based on the characteristics of speckle or its distribution. There are three main classes of digital filters; sliding window, adaptive statistical based, and edge preserved patch or pixel correlation based filters. Among 25 filters that we explored, filters #1 to #4 are median filter with window sizes 3, 5, 7, and 9 pixels, respectively; filters #5 to #8 are averaging filter with window sizes 3, 5, 7, and 9 pixels, respectively; filters #9 to #12 are symmetric nearest neighborhood (SNN) (16) with window sizes 3, 5, 7, and 9 pixels, respectively; filters #13 and #14 are Kuwahara (21) with window sizes of 3 and 5 pixels, respectively; filters #15 to #18 are adaptive Wiener filter with window sizes 3, 5, 7, and 9 pixels, respectively (22); filters #19 to #22 are Lee filter with window sizes 3, 5, 7, and 9 pixels, respectively; filter #23 is a pixel-wise NLM filter (23); filter #24 is a TV (16) filter; and filter #25 is a BM3D (24).

*Sliding window filters*: this class of filter includes mean, median and SNN, and are efficient enough to be used in real-time speckle reduction applications, such as video-rate OCT imaging. Although they effectively reduce speckle noise in the OCT image, they smooth edges in the image and create blurriness. Mean filter is a linear convolutional low-pass filter. In this filter, a pixel value is replaced by the average of its neighboring pixel values. In Median filtering a pixel value in a window, M by N pixels, is replaced by the value of the middle pixel in a vector of pixel values sorted in an ascending order (25). This nonlinear filter is more robust than the mean filter, and preserves edges more effectively. SNN, is considered as an edge preserving sliding window speckle reduction method. In SNN, initially the opposite pairs of pixels in the support are compared and replaced with the pixel value that is closest to the central pixel value (16). Each pixel value is then replaced by the average of processed pixel values in the window.

*Adaptive statistical based filters*: this class of despeckling filter, includes Kuwahara filter (11), adaptive Wiener filter, and Lee filter, and utilizes statistical features e.g., mean and variance, extracted from the image or a part of the image. Kuwahara works by dividing the support into four sub-regions (21). The central pixel is replaced with the average of the quadrants with the lowest variance. Wiener filter tailors itself to image local mean and local variance, i.e., the larger the variance, the less smoothing is applied (26). Lee filter is an adaptive filter that determines each pixel value according to the weighted sum of the center pixel value; the local statistics (mean and variance) calculated in a square kernel surrounding the pixel with a minimum mean square error (MMSE) approach (27).

*Patch or pixel correlation based filters*: this class of despeckling filter, includes NLM, TV and BM3D, are based on high inter- or intra-correlations among nearby pixels or patch of pixels. The NLM filter algorithm changes the value of the target pixel by taking the average value of all or selected pixels in the image and weighting them based on their similarity to the target pixel. NLM filters are known to preserve the textures (23). TV filters are based on an edge preserving total variation regularization process that relies on the fact that signals with excessive detail have high total variation, implying a large integral of absolute gradient for the signal (28). TV provides a close match to the ground truth image. TV efficiently suppresses the noise while preserving the image details. BM3D is a collaborative filtering method designed by considering that image has a locally sparse representation in the transform domain (24). The procedure begins with grouping similar image patches into three dimensional (3D) groups. Then a 3D linear transformation is applied on the image and a shrinkage procedure is performed. Following this process, an inverse transformation is applied on spectrum coefficients, and then, combining the patches results in an estimation of the ground truth image. Next, a Wiener filter is used to form the final denoised image (29).

## 2.2 Image Optical and Textural Features

To quantify tissue properties, 27 features including 26 texture features (30, 31) and 1 optical property are extracted from OCT images. For each image, 6 first-order statistical features including mean, variance, standard deviation, skewness, median and kurtosis and the entropy are calculated. Twenty features from the gray-level co-occurrence matrix (GLCM) (32) i.e., homogeneity, contrast, energy, entropy and correlation in four directions, 0°, 45°, 90° and 135°, are calculated as textures (33). The optical property calculated for the OCT image is the attenuation coefficient. We used Vermeers' method (34) to calculate the attenuation coefficient for each pixel in the OCT intensity image. We utilized principal component analysis (PCA) algorithm to reduce the dimension of the features (from 27 to 5). PCA analysis is a projection-based method that reduces the computational complexity through construction of orthogonal principal components while the most important variance is retained (35).

## 2.3 Image Quality Assessment Measures

The performance of the filtering methods is assessed using well established objective assessment measures, including signal to noise ratio (SNR), contrast-to-noise ratio (CNR) (8, 36-38), equivalent number of looks (ENL), mean structural similarity (MSSIM) index and edge preservation index (EPI) measures (9, 39, 40). SNR, compares the signal of an object in the OCT image to its background noise. CNR is a measure of the signal fluctuations to the noise. The definition of SNR and CNR are given in equations (1) and (2), respectively.

$$SNR = 10 log_{10} \left( \frac{\max(I_{lin}^2)}{\sigma_{lin}^2} \right) \quad (1)$$

$$CNR = \frac{1}{R} \left( \sum_{r=1}^{R} \frac{(\mu_r - \mu_b)}{\sqrt{\sigma_r^2 + \sigma_b^2}} \right) \quad (2)$$

where, $\max(I_{lin}^2)$ the linear magnitude image is the maximum of squared intensity pixel values in a homogenous region of interest (ROI), $\sigma_{lin}^2$ is the variance of $I_{lin}$ in a background noise region $\mu_b$ and $\sigma_b$ represent the mean and variance of the background, $\mu_r$ and $\sigma_r^2$ represent the mean and variance of the $r^{th}$ ROI (36). ENL is a measure of smoothness in a homogeneous ROI, and can be calculated by equation (3).

$$ENL = \frac{1}{H}\sum_{h}^{H} \mu_h^2 \sigma_h^2 \qquad (3)$$

where, $\mu_h^2$, $\sigma_h^2$ are the mean and variance of homogeneous ROIs (H). The MSSIM index quantifies image quality referring to its structural similarities and is based on local statistic calculations, equation (4).

$$MSSIM = \frac{1}{MN}\sum_{i=1}^{M}\sum_{j=1}^{N} \frac{(2\mu_{I_{(i,j)}}\mu_{\hat{I}_{(i,j)}} + C_1)(2\sigma_{I_{(i,j)}}\sigma_{\hat{I}_{(i,j)}} + C_2)}{(\mu^2{}_{I_{(i,j)}} + \mu^2{}_{\hat{I}_{(i,j)}} + C_1)(\sigma^2{}_{I_{(i,j)}} + \sigma^2{}_{\hat{I}_{(i,j)}} + C_2)} \qquad (4)$$

where, M and N are the size of the image in transverse directions, $I$ is the original image, and $\hat{I}$ is the despeckled image. $I_{(i,j)}$ and $\hat{I}_{(i,j)}$ are derived by convolving the original and despeckled images with a symmetric Gaussian kernel with window size 11 to calculate their local variance and mean, i.e., $\sigma_{I_{(i,j)}}$, $\sigma_{\hat{I}_{(i,j)}}$, $\mu_{I_{(i,j)}}$ and $\mu_{\hat{I}_{(i,j)}}$. $C_1$ and $C_2$ are constants: $C_1 = 6.502$, $C_2 = 58.522$ (40). The ground truth image is generated by averaging 170 (can be considered as spatial compounding method (41) since the images are taken from slightly misaligned samples due to the imperfect optical scanners) B-scan images to calculate MSSIM index.

EPI is a correlation-based method that shows how the edges in the image degrades.

$$EPI = \frac{\sum_{i=1}^{M}\sum_{j=1}^{N}(\Delta I_{(i,j)} - \mu_{\Delta I_{(i,j)}})(\Delta \hat{I}_{(i,j)} - \mu_{\hat{I}_{(i,j)}})}{\sqrt{\sum_{i=1}^{M}\sum_{j=1}^{N}(\Delta I_{(i,j)} - \mu_{\Delta I_{(i,j)}})(\Delta \hat{I}_{(i,j)} - \mu_{\hat{I}_{(i,j)}})}} \qquad (5)$$

where $I$ indicates the original image, $\hat{I}$ is the despeckled image, $\Delta I$ is an edge detected image with a Laplacian operator, $\mu$ and $\sigma$ are the mean and variance of the image, $\mu_b$ and $\sigma_b^2$ indicate the mean and variance of the background.

*2.4 Learnable Despeckling Framework (LDF)*

The architecture of LDF includes two main parts: (i) an autoencoder neural network, (ii) filter classifier. In section 2.4.1, we explain the architecture of the autoencoder utilized to learn the FOM. Then in section 2.4.2, we describe the architecture of the classifier that can be trained based on FOM to predict the most effective despeckling filter.

*2.4.1 Autoencoder Architecture*

A figure of merit is defined as a single representative measurement to assess the performance of each filter. In this study, we define the FOM based on a set of five OCT quality measures including SNR, CNR, ENL, EPI and MSSIM. The goal is to find an FOM which is the best representation of the quality of the image. Here, we utilized an autoencoder neural network with 3 layers for unsupervised learning of FOM. The structure of the autoencoder is illustrated in Fig. 1. As it is shown, in the layer 1, the input neurons to the network are SNR, CNR, ENL, EPI, MSSIM, and a bias neuron. Layer 2 includes one neuron to estimate the FOM, and a biased neuron. Autoencoders work well if the initial weights are close to a good solution (42). In this experiment, the initial weights are equal to 1, following the work in (8).

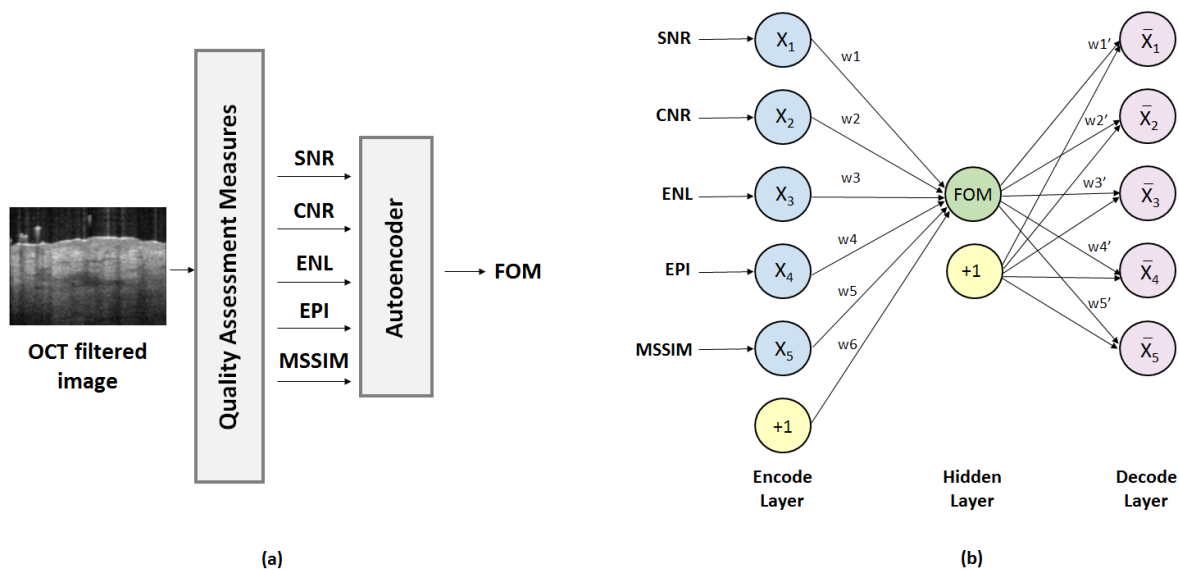

**Fig. 1** (a) Block diagram of FOM (figure of merit) calculation, (b) the structure of autoencoder, $X_1$ to $X_5$ are nodes of the encode layer, w1 to w6 and w1' to w5' are the weigh parameters, $\bar{X}_1$ to $\bar{X}_5$ are nodes of the decode layer. FOM is node of the hidden layer.

The Autoencoder is trained to calculate the FOM by utilizing the quality assessments measures of the filtered images. In this experiment, the final weights of the Encoder layer were calculated as [w1 w2 w3 w4 w5] = [ 0.1237, 0.2387, 0.0296, 0.1987, 0.4093], where the vector [w1 w2 w3 w4

w5] is corresponded to [SNR, CNR, ENL, EPI, MSSIM]. Based on this experiment, one can conclude that the MSSIM is has the higher effect on the FOM, whilst the ENL has a lower effect.

*2.4.2 Classifier Architecture*

We used the FOM measure to classify the filters. In this study, we utilized an artificial neural network (ANN) as the classifier. The classifier predicts the most effective filter (the winner filter) for the given input image. The designed ANN classifier includes three layers, the input layer, the hidden layer, and the output layer. The input layer includes 5 neurons corresponding to 5 features extracted from the image (worth to mention that the number of extracted features is 27 initially, which is reduced to 5 by utilizing PCA feature selection algorithm). The hidden layer includes 10 neurons. And finally, the output layer includes 25 neurons which is equal to the number of filters in the experiment. Fig. 2, illustrates the architecture of the designed classifier. The value of each neuron in the output layer is a real number between 0 and 1 that represents the probability of the corresponding filter being the winner filter. At the end, the filter with the highest probability will be selected as the winner. Before training, features are normalized. The features are normalized once over each image, and then normalized for the ensemble of all images.

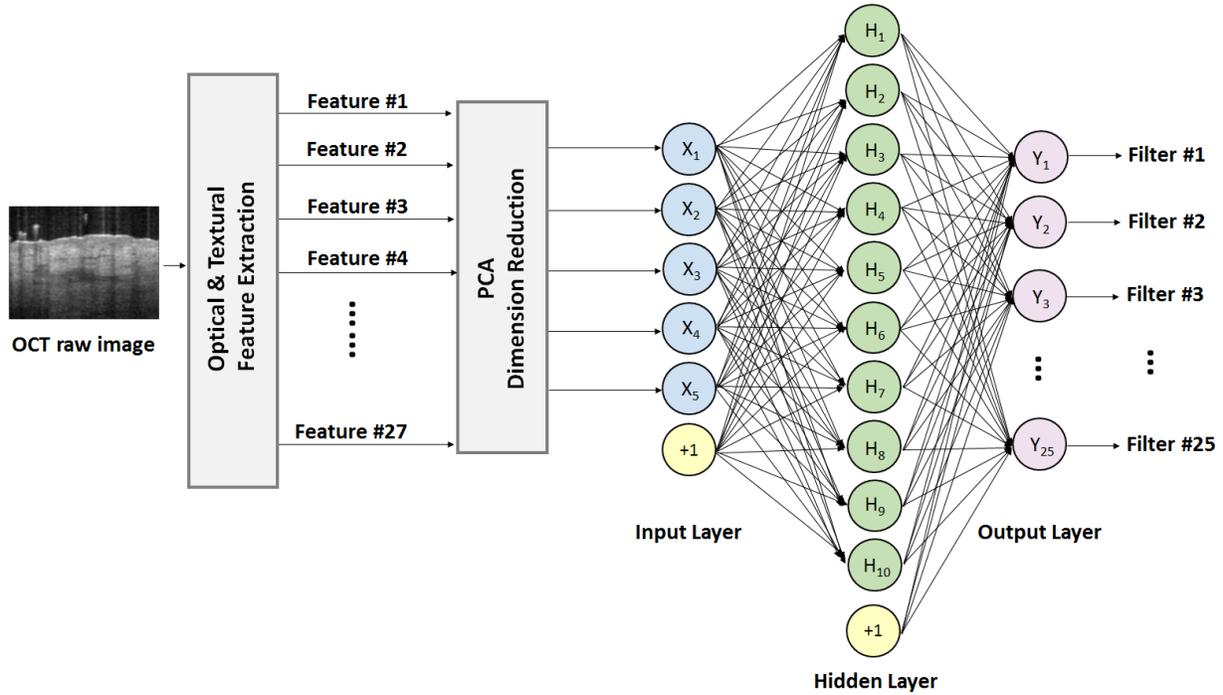

**Fig. 2** The architecture of the designed classifier. The network is trained to select the best filter for the given input OCT image. $X_1$ to $X_5$ are input nodes, $H_1$ to $H_{10}$ are hidden nodes, and $Y_1$ to $Y_{25}$ are output nodes.

The flow chart of the LDF (including classifier's training and testing steps) is given in Fig. 3. As it is shown in Fig. 3, first we are provided with a set of OCT row skin images, where in the beginning we process all the images using intensity linearization and normalization. The processed images are then passed through two parallel channels, A and B. In channel A, we extract the optical and textural features of the images and then we convert the set of the correlated features to a smaller set of linearly uncorrelated features by utilizing PCA algorithm. In channel B, we apply all the filtering methods on a selected image to create a set of filtered images. Then we calculate the FOM of each filtered image using our pretrained autoencoder. Thereafter, the filter that achieves the highest FOM is chosen as the winner filter and hence the class label of the selected image. This process repeats until all images has a class label related to them. At the end, the classifier is trained and tested using the selected features and the class labels.

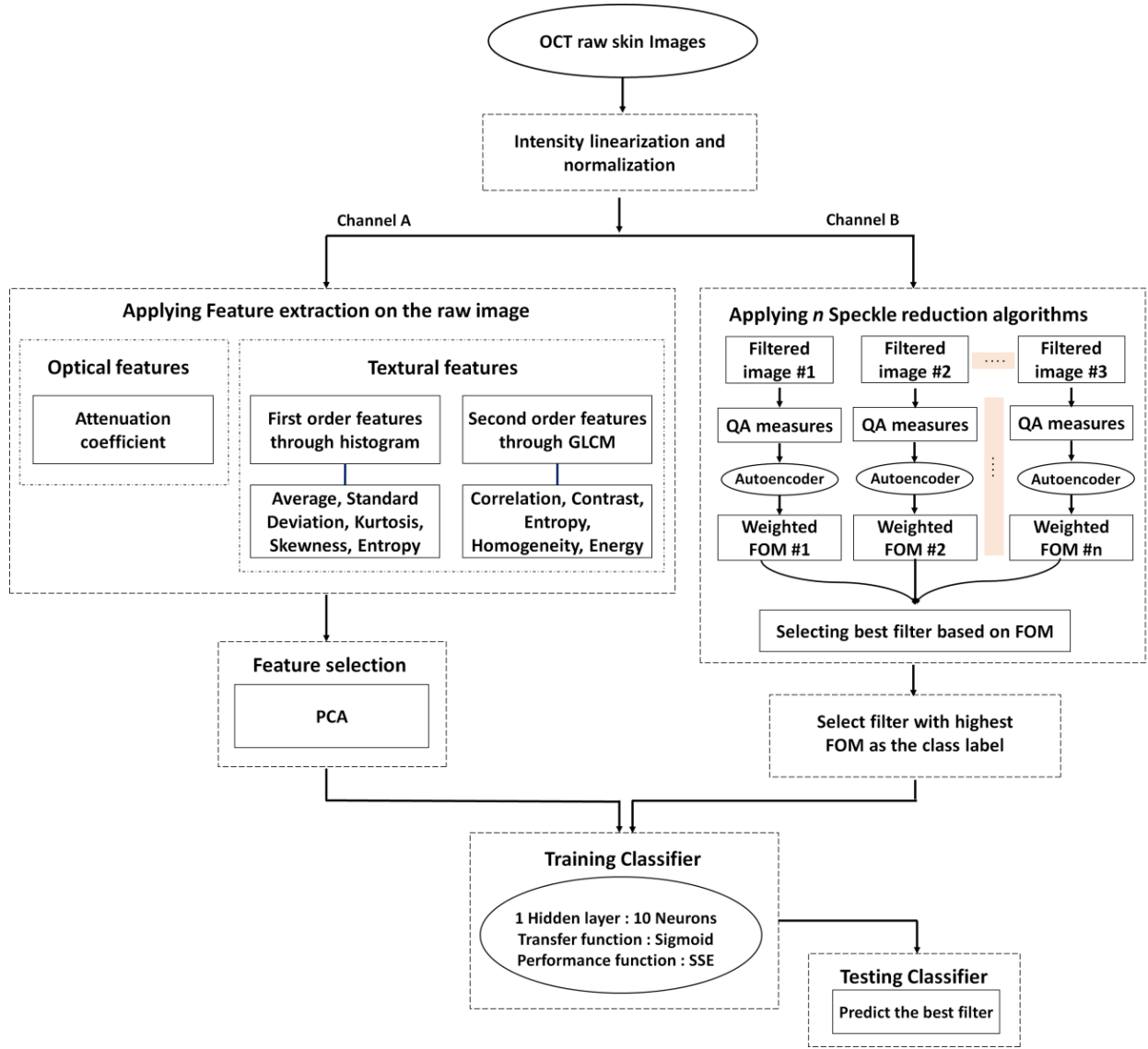

**Fig. 3** Block diagram of the LDF algorithm. OCT: optical coherence tomography, QA: quality assessment, FOM: figure of merit, and GLCM: gray level co-occurrence matrix. QA measures include SNR, CNR, ENL, MSSIM, and EPI, n is number of filters (in this study n = 25), SSE is sum square error as performance function.

In the following, we present the characteristics of OCT system used in our study.

*2.5  OCT System Specifications*

We use a multi-beam swept-source OCT system (SS-OCT) (VivoSight, Michelson Diagnostic ™ Inc., United Kingdom) for this study (43). The lateral and axial resolutions are 7.5 µm and 10 µm, respectively. The scan area of the OCT system is 6 mm (width) by 6 mm (length) by 2 mm (depth). A tunable broadband laser source with the central wavelength of 1305 +/- 15 nm, successively sweeps through the optical spectrum, leads the light to four separate interferometers and forms four consecutive confocal gates. The interference spectrum generated by the frequency sweep, over the whole bandwidth in time, is given with respect to frequency. The 10-kHz sweep is the frequency that one reflectivity profile (A-scan) is generated. A B-scan is then generated by combining several adjacent A-scans for different transversal positions of the incident beam.

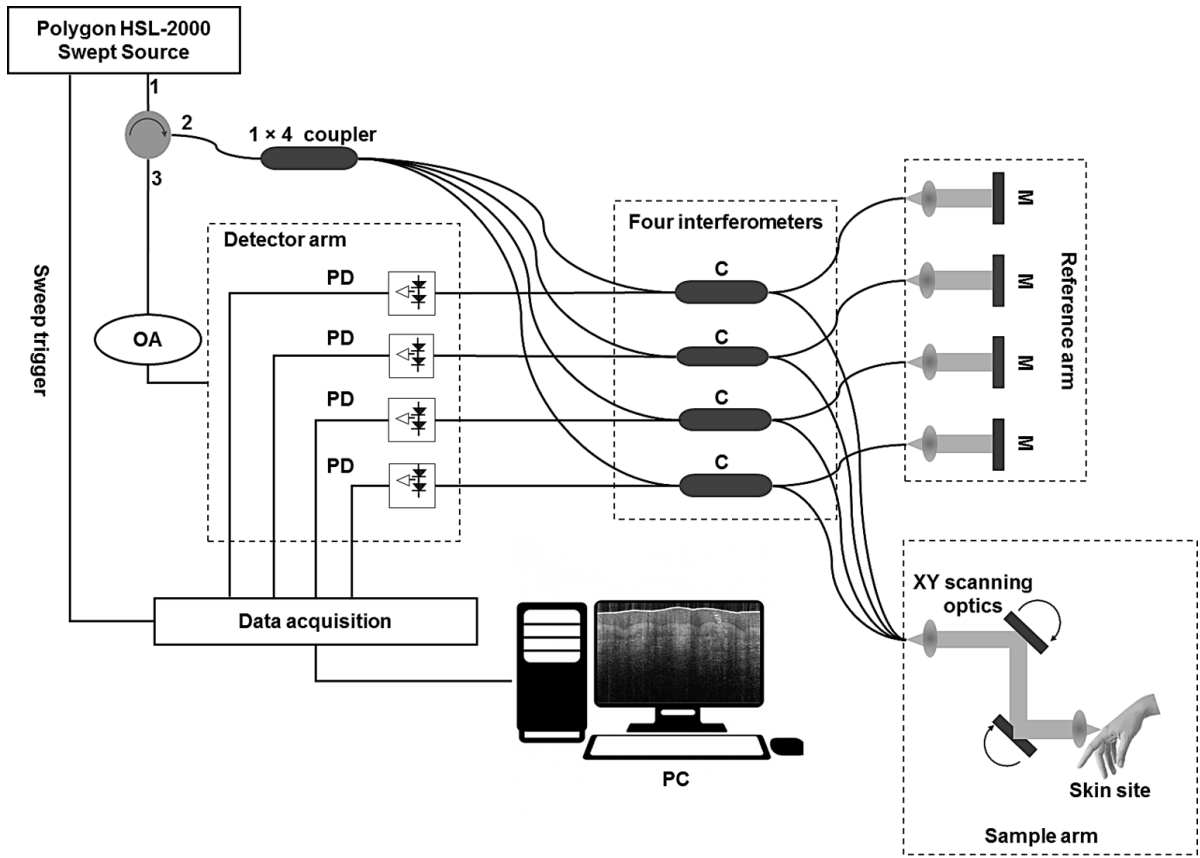

**Fig. 4** Schematic diagram of the multi-beam swept-source OCT; M: mirror, C: coupler, PD: photodetector, OA: optical attenuator, drawn by S.A.

## 3  Results and Discussion

In the following, first in subsection 3.1, we design three set of experiments, to demonstrate whether the FOM is a reliable merit and could be used as the class label in training the classifier. The experiments are performed based on the three main classes of digital filters, i.e. sliding window, adaptive statistical based, and edge preserved patch or pixel correlation based filters. Following these experiments, in subsection 3.2, we use the cross-validation technique to estimate the accuracy of the trained classifier in practice.

*3.1 FOM learning and validation*

The 27 features are computed from 25 regions of interest (ROI) in each image. For SNR and CNR calculations, 20 ROIs are utilized in images and 10 ROIs for background noise. For computation of other three quality assessment measures, the entire filtered image is used. All the 25 digital filters described in section 2, were implemented in Matlab® 2015. We used a Dell desktop computer with an Intel Core i5, 3.10 GHz CPU and 8 GB of RAM to implement the algorithms. The OCT machine is an FDA approved system for skin imaging, thus, *in vivo* skin images were collected. Images were acquired from both healthy and diseased individual's skin. OCT images of healthy tissue were taken from various body locations, to account for the variety of skin architecture found on the body. Additionally, OCT images of diseased tissue were collected, including non-melanoma skin cancer, psoriasis and acne. The imaging was performed in Oakwood Clinic, Dearborn, Michigan, USA. The institutional review board at Wayne State University (Independent Investigational Review Board, Detroit, MI) approved the study protocol. For classifier learning, the number of inputs are 285×25 images, where 285 is the number of images and 25 is the number of filters. Based on a five-fold cross validation method, out of 285×25 OCT images, 228×25 images are used for training the classifier, and the remaining (i.e., 57×25 images) are used for test (44). In the following, three experiments are performed based on the three main classes of digital filters, i.e. sliding window, adaptive statistical based, and edge preserved patch or pixel correlation based filters.

The histogram of winner filters for 285 test sets in the sliding window filter category as well as their execution time are illustrated in Fig 5. According to the graph, the average filter with the window size 5 is the winner filter for despeckling most of the time. Median filter with the window size 5, and SNN with the window size 5 are the next two winner filters.

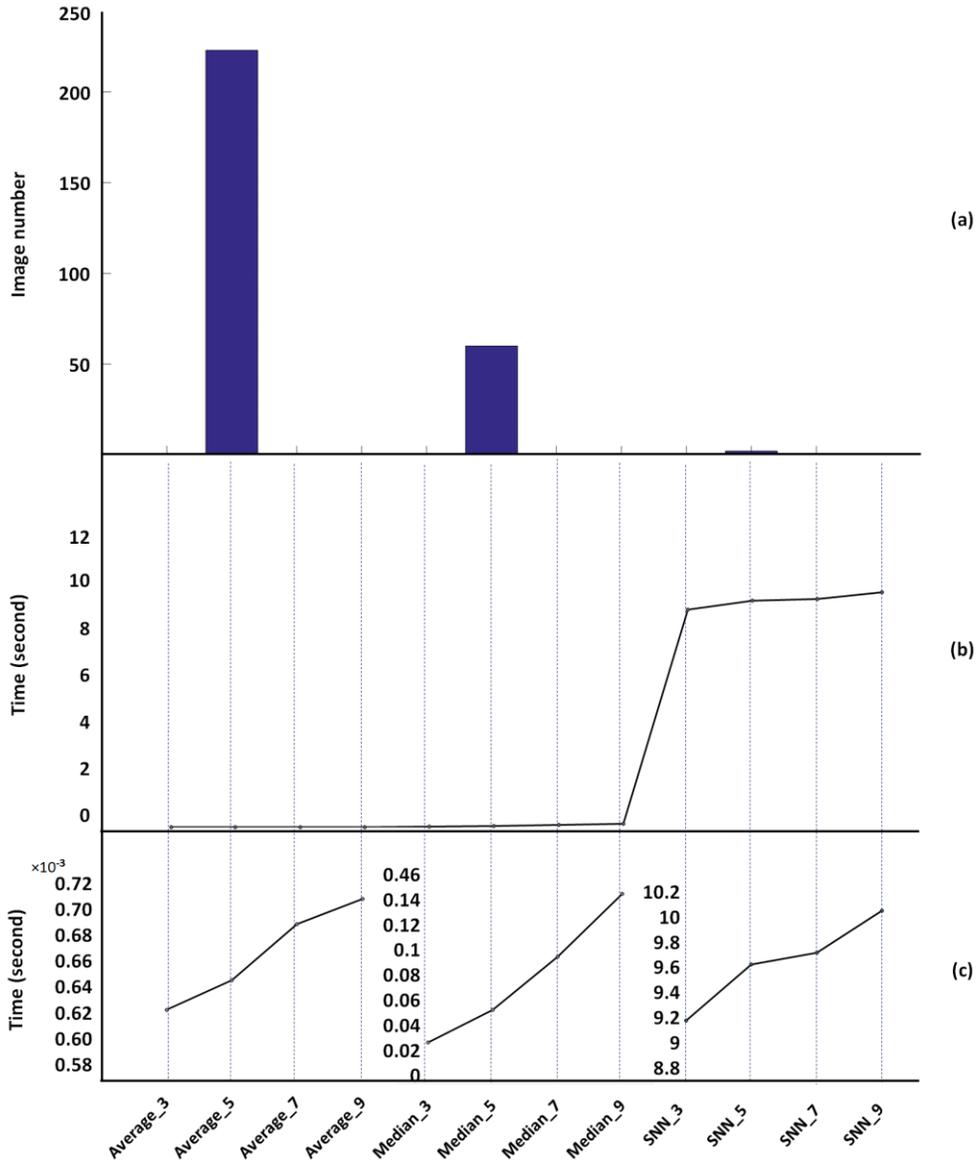

**Fig. 5** Histogram of winner filter of 285 OCT images with their corresponding execution time. Sliding window filters are used here. Histogram (a), illustrates the number of images that each filter is winning on, and histogram (b) shows the running time for each filtering method. Histogram (c) magnifies the histogram (b) in 3 distinct scales to elaborate the running time comparison between different window sizes of Average, Median and SNN filters.

In Fig. 6, three original OCT images and their despeckled ones using sliding window filters, are shown. In the case of the results in Fig. 6, the classifier has only learned the sliding window filters subgroup rather than the entire filter pool. The yellow boxes in the figure, indicate the winner filters based on FOM criterion. For the three test images, here, the average filter with the window size 5 was chosen as the winner filter. The red boxes indicate the winner filter chosen by the classifier in sliding window filtering subgroup.

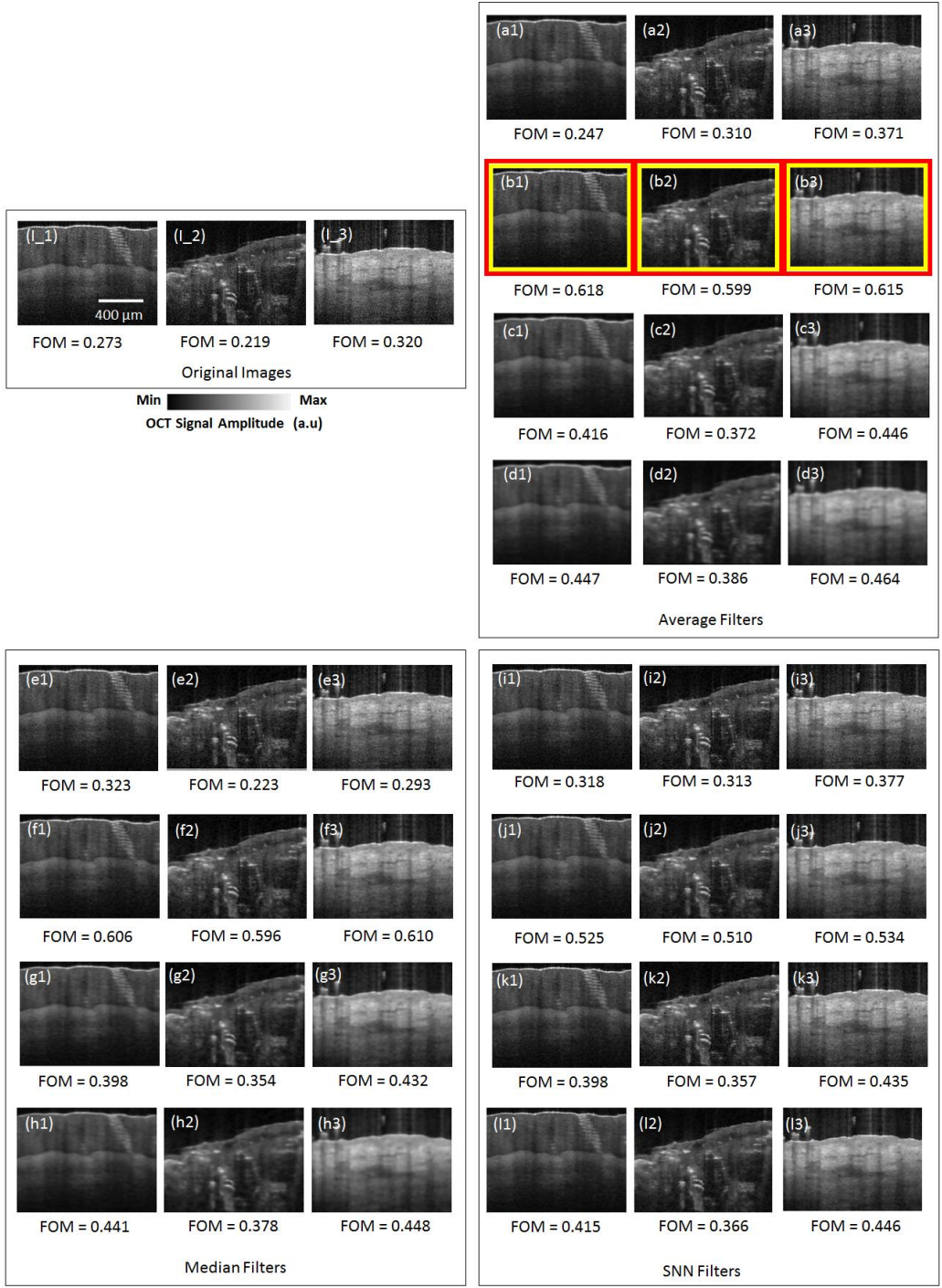

**Fig. 6** Results of despeckling using sliding window filters on three test images. Original OCT images taken from (I_1) healthy thumb of a 24-year-old male, (I_2) acne diseased outer arm of a 56-year-old female, (I_3) back of a healthy 25-year-old male, (a-d) despeckled images using averaging with window sizes 3, 5, 7, and 9, respectively, (e-h) despeckled images using median filters with window sizes 3, 5, 7, and 9, respectively, (l-l) despeckled images using SNN filters with window sizes 3, 5, 7, and 9, respectively. The yellow boxes indicate the winner filter based

on FOM measure, the red boxes indicate the winner filter chosen by the classifier in sliding window filtering subgroup.

Fig. 7, shows the histogram of winner filters in the adaptive statistical filter category for 285 OCT images as well as their execution time. According to this graph, for our image set, the Lee filter with window size 5 is chosen for all images (used in this study) when FOM is the quality assessment criterion.

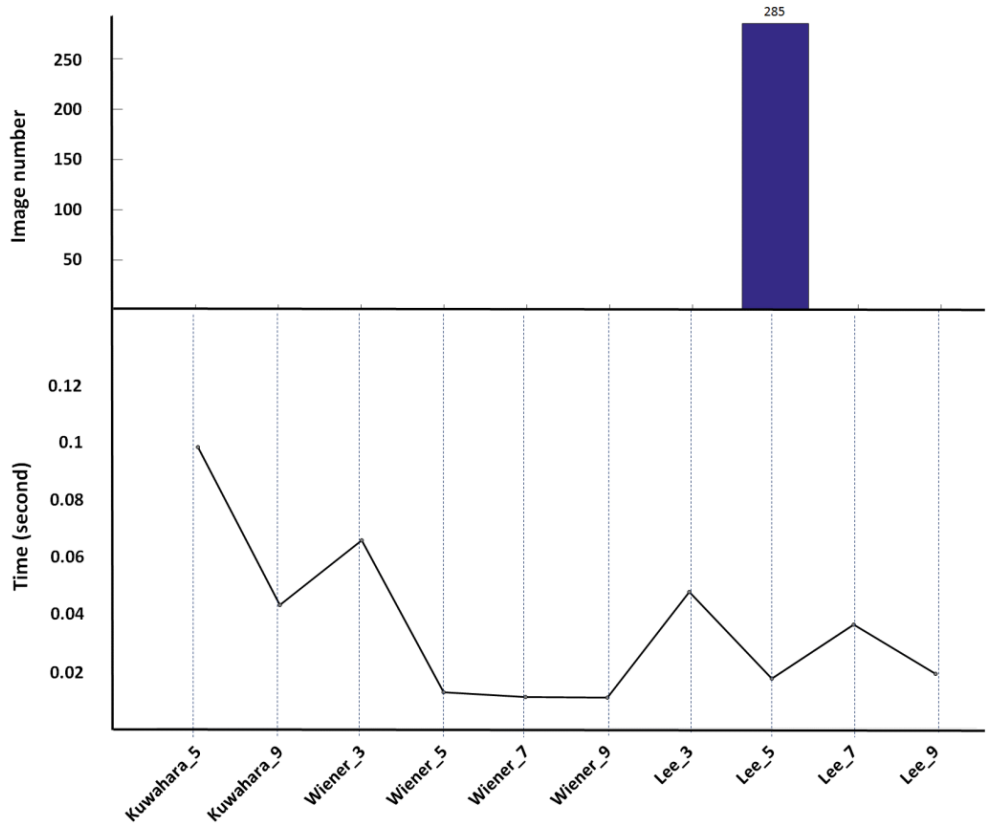

**Fig. 7** Histogram of winner filter of 285 OCT images with their corresponding execution time. Adaptive statistical filters are used here. The top histogram, illustrates the number of images that each filter is winning on, and bottom histogram shows the running time for each filtering method

In Fig. 8, original OCT images and despeckled ones using adaptive statistical filters, are shown. Here the classifier has only learned the adaptive statistical filtering subgroup. The yellow boxes in the figure, indicate the winner filters based on FOM criterion, i.e., here, Lee filter with window size 5. The red boxes indicate the winner filters chosen by the classifier in the adaptive statistical filtering subgroup.

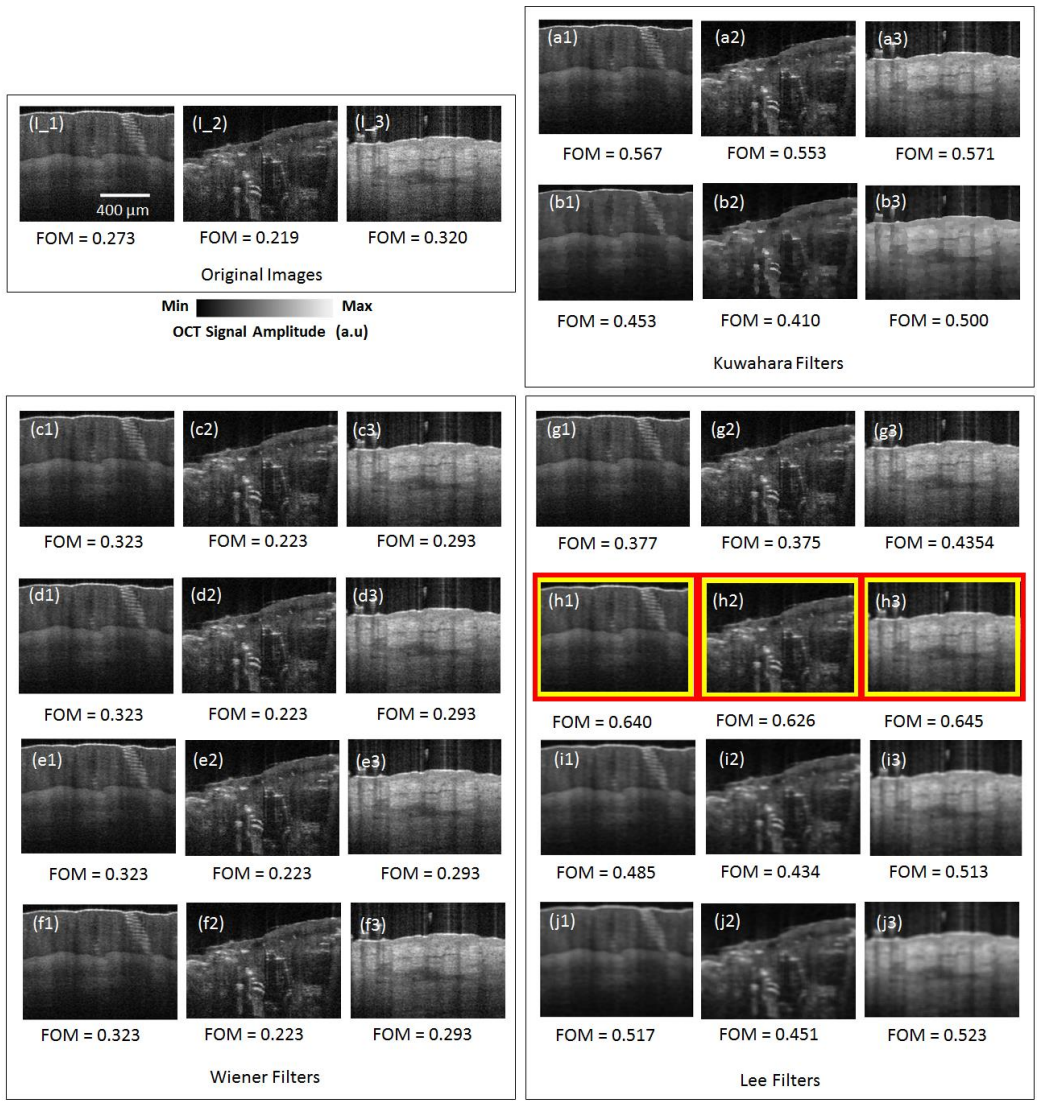

**Fig. 8** Results of despeckling using adaptive statistical filters on OCT images used in Fig. 6 (I_1-I_3). (a-b) Despeckled images using Kuwahara filter with window sizes 5, and 9, respectively, (c-f) despeckled images using Wiener filter with window sizes 3, 5, 7, and 9, respectively, (g-j) despeckled images using Lee filters with window sizes 3, 5, 7, and 9, respectively. The yellow boxes indicate the winner filter based on FOM measure, the red boxes indicate the winner filter chosen by the classifier in adaptive statistical filtering subgroup.

Fig. 9, shows the histogram of winner filters in the patch or pixel correlation filter category as well as their execution time. According to this graph, when FOM is the quality assessment criterion, in most cases BM3D filter is chosen as the winner filter. NLM and TV are the next winners, respectively.

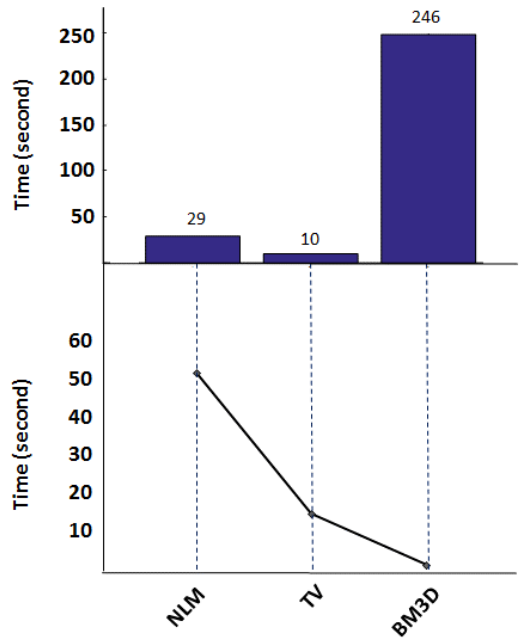

**Fig. 9** Histogram of winner filter of 285 OCT images with their corresponding execution time. Patch or pixel correlation filters are used here. The top histogram, illustrates the number of images that each filter is winning on, and bottom histogram shows the running time for each filtering method

In Fig. 10, original OCT images and despeckled ones using patch or pixel correlation filters, are shown. In this case, the classifier has only learned the patch or pixel correlation filtering subgroup. The yellow boxes in the figure, indicate the winner filters selected based on FOM criterion, i.e., BM3D and NLM. The red box indicates the winner filter chosen by classifier.

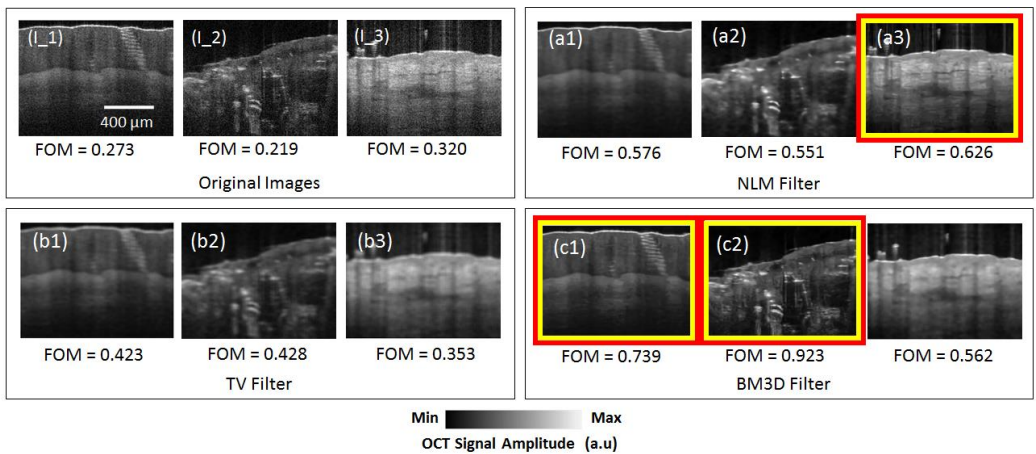

**Fig. 10** Results of despeckling using adaptive statistical filters on OCT images used in Fig. 6 (I_1-I_3). (a) Despeckled images using NLM filter, (b) despeckled images using TV filter, (c) despeckled images using BM3D

filter. The yellow boxes indicate the winner filters based on the FOM measure, the red boxes indicate the winner filter chosen by the classifier in patch or pixel correlation filters subgroup.

## 3.2 Classifier training and performance evaluation

With the above experiments and the obtained results, we demonstrated that FOM is a reliable merit and could be used as the class label in training the classifier. This was shown by comparing the winner filter chosen based on FOM (calculated regardless of image textural features and relying only on quality assessment measures) and that chosen based on the classifier (trained based on image's features). We evaluated the selection rate of filters from all three categories for the 285 images (see Fig. 11).

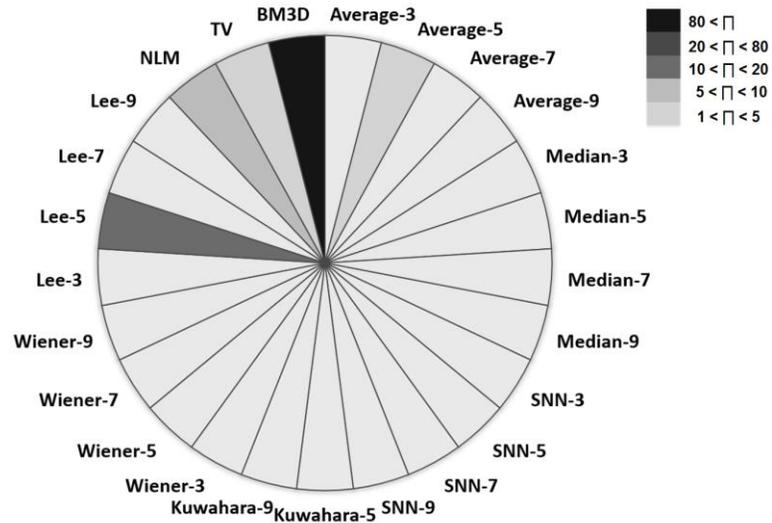

**Fig. 11** The probability map ($\prod$) of selecting each filter based on classifier's results for the 285 OCT images used in this study.

Fig. 12 (a-d) shows the results of LDF on six OCT test images with their corresponding winner filter predicted by the classifier. Fig. 12 (e-g) shows normalized FOM values for all 25 filters for the OCT images. Comparing the results of the winner filter chosen by classifier with that chosen by the normalized FOM, it is shown that the classifier can predict the winner filter with high accuracy, without having to know the result of each individual filter and only based on the features extracted from the image. In this experiment, the classifier accuracy is 97% based on a five-fold cross validation method (44).

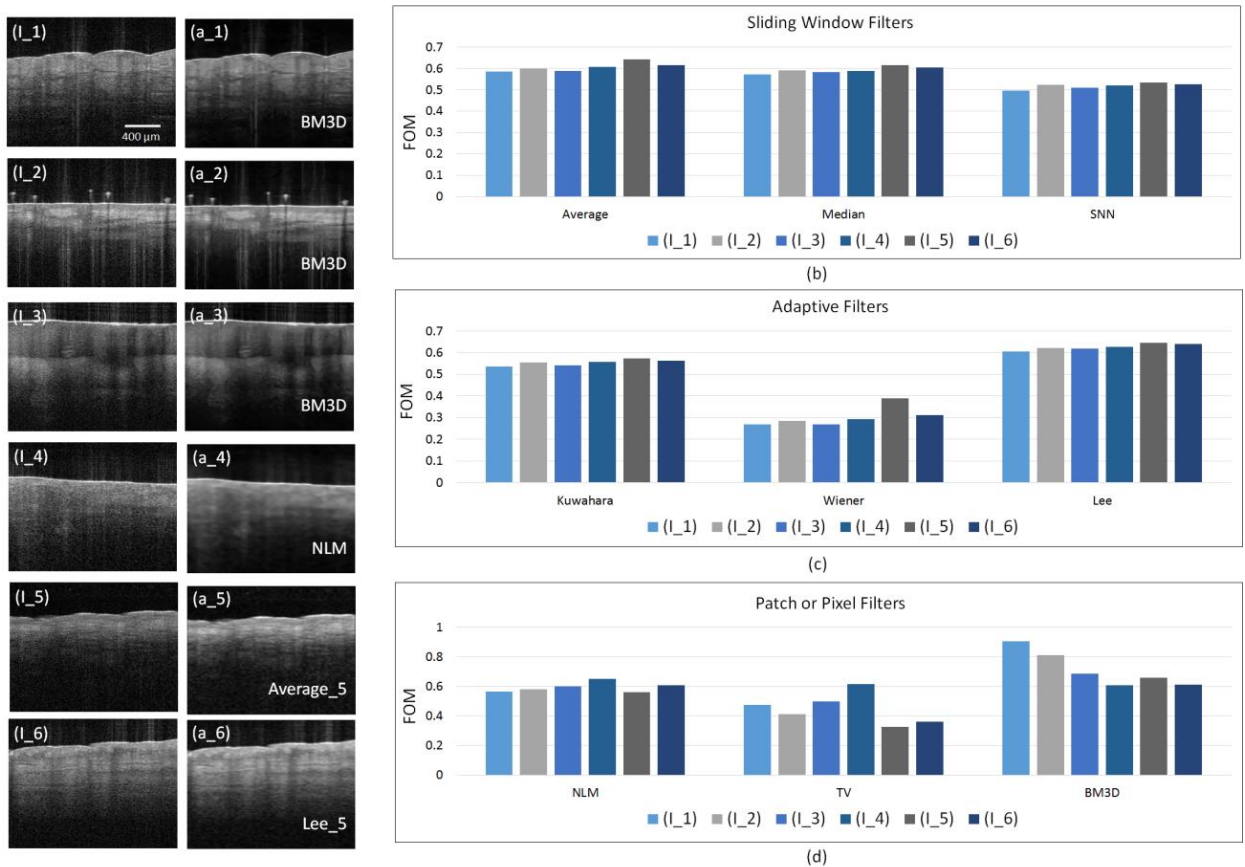

**Fig. 12** Left: Results of LDF on six OCT test images with the corresponding winner filter predicted by the classifier. (I_1-I_6) Six original OCT images, and their winner filters i.e., (a_1-a_3) despeckled images by BM3D filter as the selected optimum filter, (a_4) despeckled image by NLM filter as the selected optimum filter, (a_5) despeckled image using average filter with window sizes 5 as the selected optimum filter, (a_6) despeckled image using Lee filter with window sizes 5 as the selected optimum filter. Right: Normalized FOM for all the 25 filters for the same six OCT images for (b) sliding window filters, (c) adaptive filters, and (d) path or pixel based filters.

Referring to the execution time, we observed that even though the patch or pixel correlation category filters, filtered images most efficiently, their execution time is in the order of seconds, while the execution time of most of the sliding window filters are in millisecond range. The choice of having a better quality (based on FOM) or shorter execution time can be added to LDF criteria. Using a more diverse and a larger number of OCT images (both healthy and diseased) for training the classifier, utilizing a more efficient training algorithm along with a larger number of features, and quality assessment measures are planned as our future work.

However, several groups including our group have studied different speckle reduction methods for OCT images, finding an optimum speckle reduction filter for an image or image set, has not been comprehensively explored. We believe that LDF could help in choosing an appropriate despeckling filter based on tissue morphological, textural and optical features.

This manuscript addresses the post-processing of OCT images for biomedical applications. Speckle is a well-known artefact which affects image quality and resolution. Computer reduction of speckle has been a subject of interest since the early times of laser speckle and digital image processing, and their application to OCT has led to a large number of publications over nearly twenty years. Therefore, a wide variety of digital filters are available from the literature. We propose a procedure to learn the most appropriate speckle reduction method for one given set of images. The work is based on a large number of digital filters and a set of quality assessment measures. The learning process is trained on a set of OCT images for a given application. Each new application or acquisition procedure will be subject to its own learning stage. In a first stage, an "autoencoder" neural network is used to construct a single Figure of Merit appropriate for the given task. The We used these well-known five quality assessment measures for our study. However, by any choice of the QAS the principle should work well. Once the Figure of Merit is known, the second stage of the procedure, the so-called filter classifier, is to train a neural network to select the best filters among 25 known speckle reducing digital filters (the set of digital filters can be modified as the user sees fit) based on 5 features derived from a principal components analysis over 27 image features (with a flexibility of choice for the user). Each time, one and only one of the 25 digital filters is selected as the one that most closely matches the Figure of Merit. To illustrate and test the method, regions of interests were defined in a number of medical OCT images, including background noise. The parameters were trained on a set of 228 images and tested on further 42 images. The filter was tested in three groups: sliding window filters, adaptive statistical filters, and correlation filters. In all cases, a good match was found between the best filter selected from the Figure of Merit and the best filter selected from the "classifier".

## 4 Conclusion

In this study, we proposed a learnable framework, LDF, to reduce the speckle noise in OCT images. Initially, LDF learns a figure of merit as a single quantitative image assessment measure constructed from SNR, CNR, SNL, EPI and MSSIM using an autoencoder neural network. Then, LDF learns to decide which speckle reduction algorithm is the most effective on a given image based on image's textural and optical features. The classification accuracy of LDF is 97%. LDF is expandable, meaning that any new despeckling algorithm can easily be added to it. We believe that LDF could help in choosing an appropriate despeckling filter based on the statistical features of the tissue's OCT image.

*Disclosure*

The authors have no relevant financial interests in the manuscript and no other potential conflicts of interest to disclose.

**Saba Adabi** received her PhD in Applied Electronics from Roma Tre university in Rome, Italy in 2017. She joined Wayne State University as a visiting researcher from 2016. Her current research interests include biomedical applications of optical coherence tomography, photoacoustic imaging and ultrasound imaging.

**Caption List**

**Fig. 1** (a) Block diagram of FOM (figure of merit) calculation, (b) the structure of autoencoder, $X_1$ to $X_5$ are nodes of the encode layer, w1 to w6 and w1′ to w5′ are the weigh parameters, $\overline{X}_1$ to $\overline{X}_5$ are nodes of the decode layer. FOM is node of the hidden layer.

**Fig. 2** The architecture of the designed classifier. The network is trained to select the best filter for the given input OCT image. $X_1$ to $X_5$ are input nodes, $H_1$ to $H_{10}$ are hidden nodes, and $Y_1$ to $Y_{25}$ are output nodes.

**Fig. 3** Block diagram of the LDF algorithm. OCT: optical coherence tomography, QA: quality assessment, FOM: figure of merit, and GLCM: gray level co-occurrence matrix. QA measures include SNR, CNR, ENL, MSSIM, and EPI, n is number of filters (in this study n = 25), SSE is sum square error as performance function.

**Fig. 4** Schematic diagram of the multi-beam swept-source OCT; M: mirror, C: coupler, PD: photodetector, OA: optical attenuator.

**Fig. 5** Histogram of winner filter of 285 OCT images with their corresponding execution time. Sliding window filters are used here.

**Fig. 6** Results of despeckling using sliding window filters on three test images. Original OCT images taken from (I_1) healthy thumb of a 24-year-old male, (I_2) acne diseased outer arm of a 56-year-old female, (I_3) back of a healthy 25-year-old male, (a-d) despeckled images using averaging with window sizes 3, 5, 7, and 9, respectively, (e-h) despeckled images using median filters with window sizes 3, 5, 7, and 9, respectively, (l-l) despeckled images using SNN filters with window sizes 3, 5, 7, and 9, respectively. The yellow boxes indicate the winner filter based on FOM measure, the red boxes indicate the winner filter chosen by the classifier in sliding window filtering subgroup.

**Fig. 7** Histogram of winner filter of 285 OCT images with their corresponding execution time. Adaptive statistical filters are used here.

**Fig. 8** Results of despeckling using adaptive statistical filters on OCT images used in Fig. 6 (I_1-I_3). (a-b) Despeckled images using Kuwahara filter with window sizes 5, and 9, respectively, (c-f) despeckled images using Wiener filter with window sizes 3, 5, 7, and 9, respectively, (g-j) despeckled images using Lee filters with window sizes 3, 5, 7, and 9, respectively. The yellow boxes indicate the winner filter based on FOM measure, the red boxes indicate the winner filter chosen by the classifier in adaptive statistical filtering subgroup.

**Fig. 9** Histogram of winner filter of 285 OCT images with their corresponding execution time. Patch or pixel correlation filters are used here.

**Fig. 10** Results of despeckling using adaptive statistical filters on OCT images used in Fig. 6 (I_1-I_3). (a) Despeckled images using NLM filter, (b) despeckled images using TV filter, (c) despeckled images using BM3D filter. The yellow boxes indicate the winner filters based on the FOM measure, the red boxes indicate the winner filter chosen by the classifier in patch or pixel correlation filters subgroup.

**Fig. 11** The probability map ($\prod$) of selecting each filter based on classifier's results for the 285 OCT images used in this study.

**Fig. 12** Left: Results of LDF on six OCT test images with the corresponding winner filter predicted by the classifier. (I_1-I_6) Six original OCT images, and their winner filters i.e., (a_1-a_3) despeckled images by BM3D filter as the selected optimum filter, (a_4) despeckled image by NLM filter as the selected optimum filter, (a_5) despeckled image using average filter with window sizes 5 as the selected optimum filter, (a_6) despeckled image using Lee filter with window sizes 5 as the selected optimum filter. Right: Normalized FOM for all the 25 filters for the same six OCT images for (b) sliding window filters, (c) adaptive filters, and (d) path or pixel based filters.